# Astrobiology

## THE FAR FUTURE OF EXOPLANET DIRECT CHARACTERISATION

| | |
|---|---|
| Journal: | *Astrobiology* |
| Manuscript ID: | draft |
| Manuscript Type: | Education Articles |
| Date Submitted by the Author: | |
| Complete List of Authors: | Lammer, Helmut; Austrian Academy of Sciences, Space Reseaerch Institute |
| Keyword: | Atmospheric Compositions, Bioastronomy, Biomarkers, Biosignatures, Exobiology |
| | |

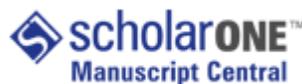

# THE FAR FUTURE OF EXOPLANET DIRECT CHARACTERISATION

**Jean Schneider[1], Alain Léger[2], Malcolm Fridlund[3], Glenn J. White[4,5], Carlos Eiroa[6], Thomas Henning[7], Tom Herbst[7], Helmut Lammer[8], René Liseau[9], Francesco Paresce[10], Alan Penny[5,11], Andreas Quirrenbach[12], Huub Röttgering[13], Franck Selsis[14], Charles Beichman[15], William Danchi[16], Lisa Kaltenegger[17], Jonathan Lunine[18], Daphne Stam[19], Giovanna Tinetti[20]**

[1]Observatoire de Paris-Meudon, LUTH, Meudon, France
[2]IAS, Universite Paris-Sud, Orsay, Paris, France
[3]RSSD, ESA, ESTEC, Noordwijk, The Netherlands
[4]The Open University, Milton Keynes, England
[5]Space Science & Technology Department, CCLRC Rutherford Appleton Laboratory, Chilton, Didcot, Oxfordshire, England
[6]Universidad Autonoma de Madrid, Madrid, Spain
[7]Max-Planck Institut für Astronomie, Heidelberg, Germany
[8]Space Research Institute, Austrian Academy of Sciences, Graz, Austria
[9]Dept. of Radio and Space Science, Chalmers University of Technology, Onsala, Sweden
[10]INAF, Via del Parco Mellini 84, I-00136, Rome, Italy
[11]Royal Observatory Edinburgh, Blackford Hill, Scotland
[12]Landessternwarte, Heidelberg, Germany
[13]Leiden Observatory, Leiden, The Netherlands
[14]Bordeaux Observatory, Floirac, France
[15]NASA ExoPlanet Science Institute, California Inst. Of Technology/JPL, USA
[16]Goddard Space Flight Center, Greenbelt, MD, USA
[17]Harvard Smithsonian Center for Astrophysics, Cambridge, MA, USA
[18]Lunar and Planetary Laboratory, University of Arizona, USA
[19]SRON, Netherlands Institute for Space Research, Utrecht, The Netherlands
[20]Department of Physics and Astronomy, University College London, London, UK

Corresponding Author:
Jean Schneider
E-mail: Jean.Schneider@obspm.fr
Observatoire de Paris-Meudon,
LUTH, Meudon, France

**Submitted to ASTROBIOLOGY**

# THE FAR FUTURE OF EXOPLANET DIRECT CHARACTERISATION

## ABSTRACT

In this outlook we describe what could be the next steps of the direct characterization of habitable exoplanets after first the medium and large mission projects and investigate the benefits of the spectroscopic and direct imaging approaches. We show that after third and fourth generation missions foreseeable for the next 100 years, we will face a very long era before being able to see directly the morphology of extrasolar organisms.

## 1.  INVESTIGATING MORE EXOPLANETS OR HABITATS

The future of exoplanetology is *a priori bright* since we already know that at least 30 % of main sequence stars have one or several super-Earth companions (Mayor 2008). It is likely that there will be two generations of space missions for the direct characterization of exoplanets in the next 15-20 years:

> ➢ a first generation with a 1.5-2 m class coronagraph suited for giant planets and near-by super-Earths (e. g. Schneider et al. 2009)

followed by:

> ➤ a second generation consisting of an interferometer (Cockel et al. 2009, Lawson et
>   al. 2009), an external occulter (Glassman et al 2009), a large 8-m class
>   coronagraph (visible – Shaklan and Levine 2008), a Fresnel Interferometric
>   Imager (Koechlin et al 2009) or a 20-m segmented coronagraph (super-JWST) for
>   the NIR (Lillie et al. 2001).

In parallel there will likely be coronagraphic cameras on ground based Extremely Large
Telescopes (like the EPICS camera - Kasper et al 2008). Here we try to anticipate what
should come next and in particular a third generation mission and beyond.

There are essentially two directions to go: investigating more planets and making a
deeper characterization of exoplanets for which a candidate biomarker has been found
and securing these biomarkers. We address these possibilities from the scientific
objectives point of view rather than from the mission design point of view. We will limit
ourselves to planets larger than 0.5 Earth radius although smaller planets may considered
as habitats but may not evolve in Earth-type habitable planets (Lammer et al. 2009).

More exoplanets at the same distance from the Earth than those detected by the two first
generation missions do not require another type of mission. To detect more habitable
exoplanets implies to observe more distant stars (hereafter "more distant" means further
than 50 pc), or telluric-like moons of know giant planets in the habitable zone of their
nearby star (hereafter "nearby" means closer than 20 pc). In both cases, what is required
is an increase in angular resolution, either to separate distant planets from their parent star or to separate a telluric companion from its parent giant planet. At 50 pc the baseline required to separate a planet at 1 AU from the star is $B = 12$ m at 600 nm[1] (resp. $B = 200$ m at 10 micron). But for distant planets angular resolution is not sufficient. The collecting area must also scale as $D^2$ where $D$ is the distance of the planetary system. For a single aperture this condition is automatically fulfilled, for an interferometer it is an extra constraint.

## 2. DEEPER CHARACTERIZATION OF MOST INTERESTING PLANETS

### 2. 1. Spectroscopic and polarimetric approach

One can search for weaker lines (like $CO_2$ around 9.3 and 10.5 micron in the thermal emission spectrum - Fig. 1), for more narrow lines and for the detailed spectral line shape thanks to higher spectral resolution. The latter case is well illustrated by the double peak of the $O_2$ band at 760 nm and of $O_3$ at 9.6 micron and the $CO_2$ central peak at 15 micron (Fig. 1). The $CO_2$ central peak is an interesting diagnostic of temperature inversion in the planet atmosphere: the central structure of the $CO_2$ band tells that the upper atmosphere is warmer than the mid-altitude regions, indicating the presence of an absorbing gas in the former region[2]. Spectroscopy of giant planets with sufficient SNR and spectral resolution will also allow measuring their radial velocity (RV). This will incidentally leave to the determination of the mass of their parent star with an unprecedented accuracy and will allow detecting and measuring the mass and distance to the planet of moons by measuring the period and amplitude of the RV variation. Indeed, the amplitude of the planet RV variation for a moo at a distance $a_{Moon}$ from its planet is $K = 0.75 (M_{\text{Moon}}/M_{\text{Io}})(M_{\text{Pl}}/M_{\text{Jup}})^{-1}(a_{\text{Moon}}/a_{\text{Io}})^{-1/2}$ km s$^{-1}$. The polarimetric approach will improve the knowledge of clouds, surface, rings etc (Stam 2008).

---

[1] We assume that planets must be at an angular distance larger than 2 $\lambda/B$ to be observable by a coronagraph:

[2] For a strongly absorbing gas, the intensity of the emission, at a given wavelength, is essentially that of a blackbody at the temperature of the atmosphere at the altitude where the optical depth $\tau$ from outside the background is unity. Consequently, the emission at the center of a band comes from regions at higher altitude than that in the band's wings, revealing the temperature of these regions.

### 2.   2. Direct imaging approach

By progressively increasing the angular resolution, one will be able to:

➢ Detect habitable moons of giant planets by separating the moon's image from its parent planet. To separate a moon distant by the Io-Jupiter distance (0.003 AU) from its parent giant planet at 10 AU requires a baseline $B = 400$ m at 600 nm (resp. $B = 7$ km at 10 micron).

➢ Improve the transit spectroscopy of transiting planets (Schneider 2000). With a baseline $B = 645$ m at 600 nm on can isolate a pixel with a size $= 0.1 R_{\text{Sun}}$ on a star at 50 pc (until now there are only 5 transiting planets closer than 50 pc) and therefore improve the SNR by a factor 10.  Note that for that case no high contrast imaging is required; it would be an excellent application of the Stellar Imager project (Carpenter et al 2009).

➢ Perform astrometric detection of moons. The astrometric measurement of the displacement of the planet position due to the pull by a moon offers another way to detect companions to planets. The required baseline is $B = 150,000$ $(D/10\text{pc})(a_{\text{Moon}}/a_{\text{Io}})^{-1}(M_{\text{Pl}}/M_{\text{Jup}})(M_{\text{moon}}/M_{\text{Io}})^{-1}$ km at 600 nm.

➤ Constrain the planet radius for transiting planets. The accurate astrometric measurement of the star's centroid during the transit of planets will give a measurement of the planet radius independent from the photometry of transits. Indeed the position of the centroid varies during the transit with a linear amplitude $R_{Pl}^3/R_*^2$. Corresponding to a few microarsec for a Jupiter-sized planet at 10 pc. (Schneider 2000).

➤ Direct measurement of the planet radius. The knowledge of the planet radius is important since this parameter controls the surface gravity and the Jeans escape of molecules. It can be inferred indirectly from the transit depth (for the few transiting planets) and constrained, with the help of atmosphere models, from the planet flux in reflected light and thermal emission. A direct measurement is obtained with an imager with an angular resolution of say 0.3 $R_{Pl}$. For a $2R_{Earth}$ planet at 5 pc the required baseline at 600 nm is $B = 20$ km.

The (temporarily) ultimate step will be the direct imaging of surface features (oceans, continents). In this configuration one can search for the direct detection of the ocean's glint (Williams and Gaidos 2008). This approach is particularly interesting for imaging *forests* and *savannahs* in order to investigate at a moderate spectral resolution the equivalent of the "red edge" of terrestrial vegetation at 725 nm. To have a, say, 10 by 10 pixel image of a $2R_{Earth}$ planet at 5 pc, a baseline B = 70 km is required at 600 nm.

These different configurations are summarized in Table 1 for an image at 600 nm. For an image at 10 micron, the required baseline is multiplied by 17. But for direct imaging the angular resolution is not sufficient. A sufficiently large collecting area is also necessary, putting an additional constraint on sparse aperture interferometers like the "hyper-telescope" (Labeyrie 1996). To have the same SNR than for a single pixel image of a planet with a 2 m (resp. 8 m) single aperture, a total area equivalent to a single aperture of 20 m, or 900 one meter apertures, (resp. 80 m, or 6400 one meter apertures) is required for have a 10 by 10 pixel image of a planet.

In conclusion, with a few exceptions, large baselines will be required in the future to perform direct imaging and, in some cases, spectroscopic observations of exoplanets. Therefore one will inevitably be led to design large interferometers, even at short visible wavelength. An intermediate step on this pathway would be a mission like the Stellar Imager (Carpenter et al. 2009) where no additional high contrast imaging performances are required.

### 3.   OTHER STUDIES

#### 4.   1. Long term monitoring of most interesting planets

A better knowledge of what is going on at the most interesting planets will be provided by long term monitoring programs from months to years.  This programs will lead to an improved knowledge of their diurnal rotation, random cloud coverage, seasons and volcanic events. A particular application is the detection of moons by *mutual events* during a continuous photometric monitoring of the planet flux. These mutual events will reveal the presence of a moon by the shadow they project on the planet, by their disappearance in the planet shadow and by the primary and secondary transit with the parent planet (Cabrera and Schneider 2007).

### 3. 2. "Techno-Signatures"

Beyond standard biosignatures, another type of far from equilibrium signals can be seen as *techno-signatures*, i.e. spectral features not explained by complex organic chemistry, like laser emissions. In the present state of our knowledge one cannot eliminate them a priori, although we have no guiding lines to search for them.  For instance, in the present Earth atmosphere, CFC (Carbon Fluoro Compounds) gases are the result of technological chemical synthesis. Observed over interstellar distances, they would reveal to the observer the presence of a technology on our planet. The detection of their absorption spectrum on an exoplanet would require a spectral resolution of at least 100 around 10 micron (See Fig. 2). Another approach would be to detect artificially produced light (e.g. Laser light). On Earth the present total energy production is about 40 TW. This represented one millionth of Sunlight energy reflected by the whole Earth. Transposed to an exoplanet it means that artificial light produced with the same power would be lost in the background noise of the stellar light reflected by the planet. This situation can be circumvented by observing the planet in the night-side only. But then the spatial resolution should be at least 0.3 $R_{Pl}$, corresponding to a baseline $B$ of 70 km for a 2 $R_{Earth}$ planet at 5 pc. To be detected with a SNR equivalent to the detection of the reflected stellar light by the whole planet with a 1.5 telescope, the collecting area required to detect artificial light one million times fainter should be one million times larger, i.e. correspond to a single aperture with a diameter B = 1.5 km. Another type of techno-signature could go beyond the suggestion to detect artificial constructions by their transits on front of stars (Arnold 2005): this kind of detection would be improved by resolving the stellar disc.

## 4. THE VERY LONG TERM PERSPECTIVE

If we suppose that around 2020-2030 one has found a promising biomarker candidate on a nearby planet (like for instance around alpha Cen (Guedes et al. 2008). Such a discovery would trigger two kinds of projects:

➤ *Direct visualization of living organisms.* Suppose that one wants to detect directly the shape of an organism having a size of 10 meter. A spatial resolution of 1 meter would be required. Even on the putative closest exoplanet alpha Cen A/B b, the required baseline would be at 600 nm $B = 600,000$ km (almost the Sun radius). In reflected light the required collecting area to get 1 photon per year in reflected light is equivalent to a single aperture of $B = 100$ km. In addition, it this organism is moving with a speed of 1 cm s$^{-1}$ it must be detected in less than 1000 sec. To get a detection in 20 minutes with a SNR of 5, the collecting area must then correspond to an aperture $B = 3$ million km. All these numbers are unrealistic, unless laser trapped mirrors proposed by Labeyrie et al (2009) finally succeed (in their present conception there are fragilized by the solar wind).

> *Exploring nearby stars.* The possibility to explore in situ nearby stars at a speed of 0.3 c is often invoked (see for instance Bjoerk 2008). But at this kind of speed on faces the problem of shielding against cosmic rays damaging electronics and the interstellar dust threatening the whole mission.   According to Semyonov (2009), a water shell of 1 m in thickness would be a sufficient protection (but then the problem of accelerating up to 0.3 c is raised). As for the threat by interstellar dust, a 100 μ interstellar grain at 0.3 the speed of light has the same kinetic energy than a 100 tons body at 100 km/hour. No presently available technology can protect against such a threat without a spacecraft having itself a mass of hundreds of tons, in turn extremely difficult to accelerate up to 0.3 c. A way around is to have a travel velocity of only a few hundred km/sec like for the "The Project" project (Kilston 1999). But then the journey will take 10,000 years to go to alpha Cen.

Whatever the approach, it seems impossible to have a direct visual contact with living organisms on the nearest exoplanet before many centuries, at least in the framework of foreseeable physical and technological concepts, and what physics will be in 1,000 years is not reasonable to anticipate. We are thus limited by a kind of conceptual or knowledge horizon.

## CONCLUSION

For the one or two next centuries, one can reasonably anticipate that first high resolution spectroscopy and then high angular direct imaging will improve considerably our knowledge of nearby exoplanets and possible global biomarkers.  For the latter approach, large interferometers will be inevitable. The highly desirable next step would be to have a direct visual access to the morphology of hypothetical life forms on these planets. Unfortunately technological obstacles will lead to a frustrating period of many centuries before one can realize this hope and we are perhaps as far from this epoch than Epicurus was far from seeing the first *other worlds* when, 23 centuries ago, he predicted the existence of these planets (Epicurus 300 BC).

# FIGURES

**Figure 1**

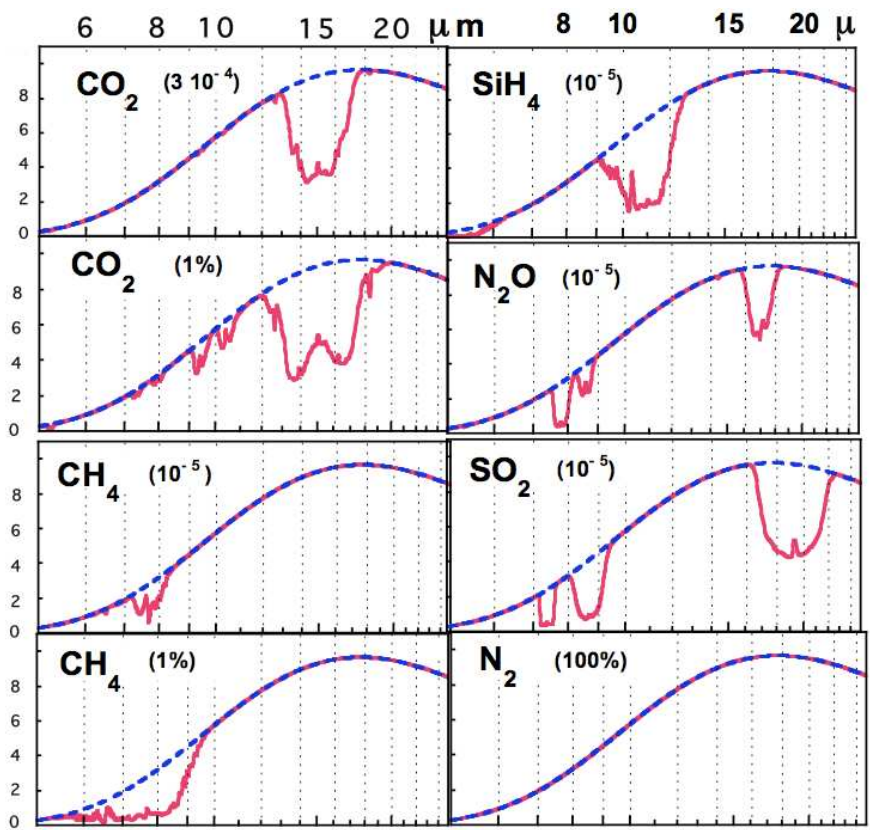

**Fig. 1:** Spectra of the thermal emission of planets somewhat similar to the Earth but with different gases in their atmospheres. The spectral resolution is $\mu\lambda/\Delta\lambda = 200$ and signal to noise ratio $\sim 100\ \sigma$. An additional non IR-active gas, e.g. $N_2$ or $O_2$, is assumed with a pressure of 1 bar. The thermal profile of the atmosphere is that of the Earth. All the spectral features present are real and show how precise are the fingerprints of gases at this spectral resolution (about 10 times larger than that of Darwin in the present version). In particular an estimate of the abundance of the gases is possible. The reader is invited to compare the cases of $CO_2$ and $CH_4$ at different concentrations (with respect to the 1 bar of inert gas). As explained in the text, the central structure of strong bands, e.g. $CO_2$, are informative about the thermal structure of the atmosphere (warm stratosphere).

**Figure 2**

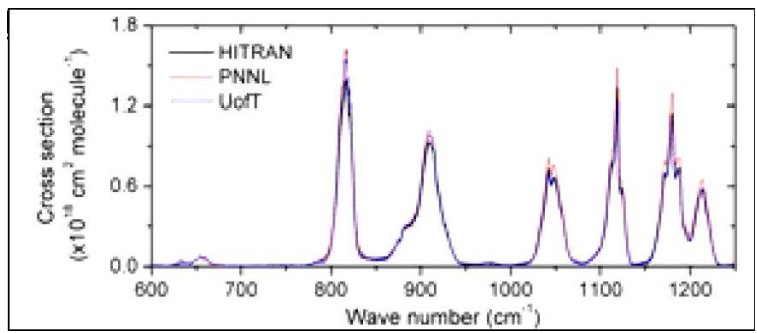

**Fig 2:** Spectrum of CFC-113 (Le Bris et al. 2006) showing good agreement between the HITRAN model, and Pacific Norhwest National Laboratory and U. of Toronto measurements.

# TABLES

| Objectives | Required baseline at 600 nm |
|---|---|
| Exo-moon image | 400 m |
| Spectroscopy of a planet transit image | 645 m |
| Astrometry of a planet transit image | 40 km |
| Direct measurement of planet size | 20 km |
| Image of continents/oceans | 70 km |

**Table 1:** Baseline distances, necessary for exo-moon investigations to direct imaging of exoplanet surface features (after Williams and Gaidos 2008).